\documentclass[12pt]{article}

\usepackage{amssymb,amsmath,amsfonts,latexsym,graphicx}
\usepackage[numbers]{natbib}
\usepackage{tabularx}
\usepackage{array}

\begin{document}

\begin{center}
\Large \bf Leveraging AI for Automatic Classification of PCOS Using Ultrasound Imaging \rm

\vspace{1cm}


\large Atharva Divekar$\,^a$, \large  Atharva Sonawane$\,^b$

\vspace{0.5cm}

\normalsize


$^a$ $^b$ Vishwakarma Institute of Information Technology, Pune, India


\vspace{5mm}


Email: {\tt hackathonpy31@gmail.com}

\vspace{6mm}

\end{center}

\abstract{The AUTO-PCOS Classification Challenge seeks to advance the diagnostic capabilities of artificial intelligence (AI) in identifying Polycystic Ovary Syndrome (PCOS) through automated classification of healthy and unhealthy ultrasound frames. This report outlines our methodology for building a robust AI pipeline utilizing transfer learning with the InceptionV3 architecture to achieve high accuracy in binary classification. Preprocessing steps ensured the dataset was optimized for training, validation, and testing, while interpretability methods like LIME and saliency maps provided valuable insights into the model's decision-making. Our approach achieved an accuracy of 90.52\%, with precision, recall, and F1-score metrics exceeding 90\% on validation data, demonstrating its efficacy. The project underscores the transformative potential of AI in healthcare, particularly in addressing diagnostic challenges like PCOS. Key findings, challenges, and recommendations for future enhancements are discussed, highlighting the pathway for creating reliable, interpretable, and scalable AI-driven medical diagnostic tools.}

\section{Introduction}\label{sec1}

The AUTO-PCOS Classification Challenge is a groundbreaking initiative aimed at exploring the capabilities of artificial intelligence (AI) in revolutionizing healthcare diagnostics, particularly for Polycystic Ovary Syndrome (PCOS). This challenge is rooted in addressing a pressing global issue: the underdiagnosis of PCOS, which affects millions of women worldwide, with profound implications on their physical, mental, and reproductive health. The overarching goal of the challenge is to foster the development of robust, scalable AI models that can automatically classify healthy and unhealthy frames extracted from ultrasound videos, thereby providing a transformative solution to a long-standing medical challenge.

PCOS is a complex endocrine disorder characterized by a combination of symptoms such as irregular menstrual cycles, excessive androgen levels, and the presence of cysts in the ovaries. Despite its prevalence, it often goes undiagnosed due to the subjective and operator-dependent nature of traditional diagnostic techniques. The AUTO-PCOS Challenge aims to bridge this gap by leveraging AI to create an automated, operator-independent diagnostic tool that can assist clinicians in making accurate and timely diagnoses. This initiative not only aligns with the growing emphasis on precision medicine but also has the potential to significantly enhance accessibility and efficiency in PCOS diagnostics, especially in resource-limited settings.

At the heart of this challenge lies the PCOSGen dataset, a meticulously curated collection of ultrasound images that forms the cornerstone of this initiative. The dataset comprises a total of 4,668 images, with 3,200 allocated for training and 1,468 for testing. These images are categorized into two distinct classes---Healthy and Unhealthy---to facilitate binary classification. Each image has been annotated by experienced gynecologists, ensuring the highest level of accuracy and reliability. The dataset has been sourced from diverse platforms, including YouTube, Kaggle, and other medical repositories, to capture a wide range of imaging variations and enhance the generalizability of the AI models developed.

\subsection{Key Characteristics of the Dataset}
\begin{itemize}
    \item \textbf{Number of Images in Each Class:}
    \begin{itemize}
        \item Healthy: \textbf{903}
        \item Unhealthy: \textbf{2297}
    \end{itemize}
    The dataset exhibits an imbalanced distribution between the two classes, as depicted in Figure~\ref{fig1:class_distribution}. This imbalance, with 903 Healthy images and 2,297 Unhealthy images, necessitates careful handling during model training to mitigate potential bias and ensure that the model performs well across both categories.

    \begin{figure}[htbp]
        \centering
        \includegraphics[width=0.6\linewidth]{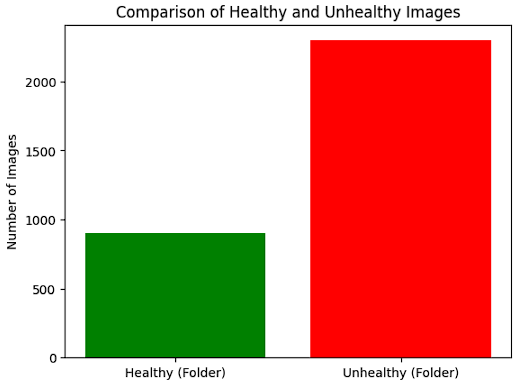}
        \caption{Bar chart illustrating the number of images in each class.}
        \label{fig1:class_distribution}
    \end{figure}
    
    \item \textbf{Pixel Intensity Analysis:}
    \begin{figure}[htbp]
        \centering
        \includegraphics[width=0.8\linewidth]{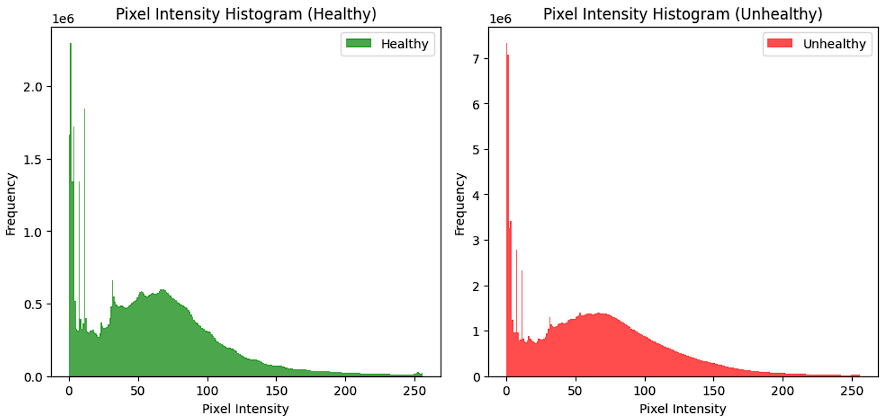}
        \caption{Pixel intensity histograms showing potential for differentiation between Healthy and Unhealthy image classes based on pixel brightness distribution.}
        \label{fig2:pixel_intensity}
    \end{figure}
    
    A histogram analysis of pixel intensity values across both classes given in Figure~\ref{fig2:pixel_intensity} reveals important distinctions between the "Healthy" and "Unhealthy" images.
    \begin{itemize}
        \item \textbf{Healthy Class:}
        The Healthy images typically exhibit a higher frequency of dark pixels with peaks near low intensity values. There is a gradual decrease in frequency as intensity increases, suggesting fewer bright pixels.
        
        \item \textbf{Unhealthy Class:}
        In contrast, the Unhealthy images show a more pronounced peak at low intensities and a broader spread in mid-range values. These images also have a slightly more gradual decline in pixel frequency as intensity increases, indicating a higher proportion of brighter pixels.
    \end{itemize}
    These differences in pixel intensity distributions suggest that pixel intensity could play a key role in distinguishing between "Healthy" and "Unhealthy" classes. For example, "Unhealthy" images tend to exhibit more variation in brightness and a higher number of brighter pixels, which could inform classification models.
    \end{itemize}

The data provided by the committee was already exceptionally clean and free from any noise or errors. Rigorous quality control measures had been implemented to ensure that the dataset was devoid of inconsistencies, making it an ideal foundation for model training. This pristine dataset allowed for seamless preprocessing, with no need for further corrections or filtering, ensuring that the focus remained on the accuracy and performance of the model. The inclusion of medically annotated labels further elevated the dataset's reliability, reinforcing its value for both academic research and clinical applications.

\section{Methods}\label{sec2}

\begin{figure}[htbp]
    \centering
        \includegraphics[width=\linewidth]{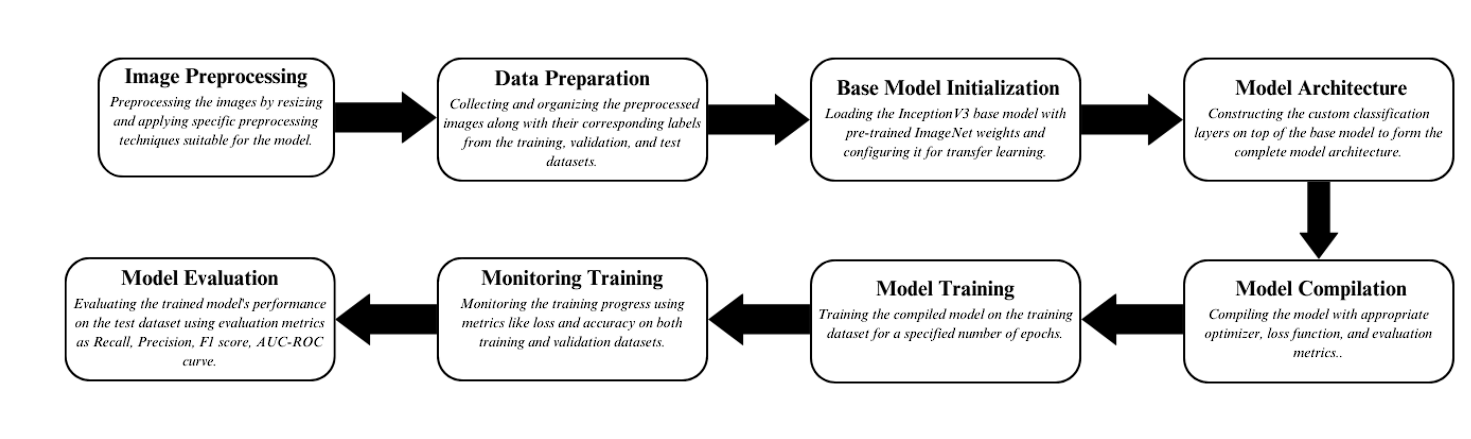}
        \caption{Deep Learning Pipeline for PCOS Image Classification. This figure illustrates the key stages of the image classification pipeline, including preprocessing, data preparation, model training, and evaluation.}
    \label{fig3:block_diagram}
\end{figure}

This section outlines the methodology employed in the submission to the AUTO PCOS challenge illustrated in Figure~\ref{fig3:block_diagram}, detailing each step of the pipeline from data preprocessing to model evaluation, along with reasoning for each step and a discussion of model experimentation.

Our methodology follows a structured approach, encompassing the following key stages:

\subsection{Image Preprocessing}
    Preprocessing is essential for preparing raw images for model analysis. It addresses variations in size, quality, and color distribution that can negatively impact performance.
    \begin{itemize}
    \item \textbf{Steps:}
    \begin{itemize}
        \item \textbf{Image Loading:} Images are loaded from directories organized by their respective labels (Healthy or Unhealthy), maintaining a clear structure for training and evaluation.
        \item \textbf{Resizing:} Images are resized to a uniform dimension of 256x256 pixels, ensuring input consistency for the neural network.
        \item \textbf{Normalization:} Pixel values are normalized using preprocessing functions specific to the InceptionV3 model. This standardizes the input data, improving training convergence by centering features around zero and scaling them appropriately.
    \end{itemize}
\end{itemize}

\subsection{Data Preparation}
    Proper data preparation is crucial for effective model learning.
    \begin{itemize}
    \item \textbf{Steps:}
    \begin{itemize}
        \item \textbf{Data Splitting:} The dataset is divided into training (70\%), validation (20\%), and testing (10\%) sets. This split prevents overfitting by ensuring the model is evaluated on unseen data.
        \item \textbf{Label Encoding:} Labels are encoded as binary values (0 for Healthy, 1 for Unhealthy), simplifying the classification task and aligning with the model's output layer.
    \end{itemize}
\end{itemize}

\subsection{Base Model Initialization}
    Utilizing a pre-trained model leverages knowledge from large datasets, improving performance on our smaller dataset through transfer learning.
    \begin{itemize}
    \item \textbf{Steps:}
    \begin{itemize}
        \item \textbf{Model Configuration:} The InceptionV3 model is initialized with pre-trained ImageNet weights, excluding the top classification layers (\texttt{include\_top=False}) to allow for customization.
        \item \textbf{Freezing Layers:} The convolutional base of InceptionV3 is initially frozen (\texttt{base\_model.trainable = False}) to preserve the pre-trained weights during the initial training phase, focusing on learning task-specific features in the added layers.
    \end{itemize}
\end{itemize}

\subsection{Model Architecture}
    A well-designed architecture is vital for effective feature extraction and classification.
    \begin{itemize}
    \item \textbf{Custom Layers (added on top of InceptionV3 base):}
    \begin{itemize}
        \item \textbf{Layer Normalization:} Normalizes activations across batches for stable training and faster convergence.
        \item \textbf{Convolutional Layer (1024 filters, ReLU activation):} Captures complex patterns in the data, with ReLU introducing non-linearity.
        \item \textbf{Max Pooling Layer:} Downsamples feature maps, reducing computational load and mitigating overfitting.
        \item \textbf{Dropout Layer (Dropout rate 0.3):} Further mitigates overfitting by randomly setting input units to zero during training.
        \item \textbf{Flattening Layer:} Converts the 2D feature maps to a 1D vector for the fully connected layers.
        \item \textbf{Dense Layer (512 units):} Further processes the extracted features.
        \item \textbf{Output Dense Layer (1 unit, Sigmoid activation):} Performs binary classification, outputting a probability between 0 and 1.
    \end{itemize}
\end{itemize}

\subsection{Model Compilation}
    Compilation configures the model for training by defining the learning process.
    \begin{itemize}
    \item \textbf{Components:}
    \begin{itemize}
        \item \textbf{Optimizer:} Adam optimizer, chosen for its efficiency with sparse gradients and adaptive learning rates.
        \item \textbf{Loss Function:} Binary cross-entropy loss, appropriate for binary classification.
        \item \textbf{Metrics:} Binary accuracy, used to monitor performance on training and validation sets.
    \end{itemize}
\end{itemize}

\subsection{Model Training}
    Training allows the model to learn patterns from the data.
    \begin{itemize}
    \item \textbf{Process:}
    \begin{itemize}
        \item The compiled model is trained on the training dataset for a specified number of epochs (e.g., 50).
        \item Validation data is used to monitor performance during training, enabling early stopping or hyperparameter adjustments.
    \end{itemize}
\end{itemize}

\subsection{Monitoring Training Progress}
    Monitoring ensures learning objectives are met and allows for adjustments during training.
    \begin{itemize}
    \item \textbf{Method:} The training \texttt{history} is used to track metrics (loss and accuracy) across epochs for both training and validation sets, allowing for visualization and analysis of training progress.
\end{itemize}

\subsection{Model Evaluation}
    Evaluation on unseen data assesses the model's generalization ability.
    \begin{itemize}
    \item \textbf{Process:} After training, the model is evaluated on the held-out test dataset using metrics such as accuracy, precision, recall, F1-score, and AUC-ROC curve.
\end{itemize}

\subsection{Model Experimentation}
    Exploring different architectures helps identify the most suitable model for the task.
    \begin{itemize}
    \item \textbf{Models Explored:} In addition to InceptionV3, we experimented with ResNet101, Vision Transformer (ViT), EfficientNet B7, and RadImagenet \cite{RadImagenet} pre-trained radiology-specific models.
    \item \textbf{Model Selection:} InceptionV3 was selected as the final model due to its superior validation accuracy on our dataset compared to the other architectures tested.
\end{itemize}

This comprehensive methodology effectively utilizes deep learning techniques for binary classification in ultrasound imaging related to PCOS diagnosis while ensuring robust evaluation and interpretability of results. By following these structured steps and conducting thorough experimentation with various models, Team Error aims to deliver a high-performing solution tailored specifically for this healthcare challenge.

\section{Results}\label{sec3}
This section presents the outcomes of the model evaluation and includes performance metrics, training and validation curves, sample predictions with LIME explanations, heatmaps, and insights into computational efficiency.

\subsection{Model Performance Metrics}

\begin{table}[h]
    \centering
    \begin{tabularx}{0.8\linewidth}{|X|X|}
        \hline
        \textbf{Metric} & \textbf{Score} \\
        \hline
        Accuracy  & 0.9052  \\
        Precision  & 0.9001  \\
        Recall  & 0.9716 \\
        F1 - Score  & 0.9345  \\
        \hline
    \end{tabularx}
    \caption{Performance Metrics of InceptionV3 Model on Validation Split of the Dataset}
    \label{tab1:performance}
\end{table}

These metrics given in Table~\ref{tab1:performance} demonstrate the model's effectiveness in classifying ultrasound images as either healthy or unhealthy.

The accuracy of the model is reported at 90.52\%, indicating that the InceptionV3 model correctly classified approximately 90.52\% of the test images. This high level of accuracy suggests that the model is proficient in distinguishing between healthy and unhealthy ultrasound images, making it a reliable tool for diagnosing conditions such as Polycystic Ovary Syndrome (PCOS).

The precision score of 90.01\% indicates that when the model predicts an image to be "unhealthy," it is correct about 90.01\% of the time. High precision is particularly important in medical diagnostics, as it minimizes the risk of false positives, which can lead to unnecessary anxiety and further testing for patients.

The recall score of 97.16\% demonstrates that the model successfully identifies 97.16\% of actual unhealthy cases. This metric highlights the model's ability to detect positive instances (unhealthy images) accurately, which is crucial in medical contexts where missing a diagnosis can have significant consequences.

The F1 score of 93.45\% provides a balance between precision and recall, offering a single metric that reflects both aspects of performance. The F1 score is particularly useful when dealing with imbalanced datasets, as it ensures that both false positives and false negatives are considered in evaluating model performance.

The combination of high accuracy, precision, recall, and F1 score indicates that the InceptionV3 model is not only effective but also reliable for classifying ultrasound images related to PCOS diagnosis. These results suggest that the model can be utilized confidently in clinical settings to assist healthcare professionals in making informed decisions based on ultrasound imaging.

\subsection{Comparion with Baseline Models}

\begin{table}[h]
    \centering
    \begin{tabular}{|l|c|c|c|c|}
        \hline
        \textbf{Model} & \textbf{Accuracy} & \textbf{Precision} & \textbf{Recall} & \textbf{F1 Score} \\
        \hline
        \textbf{InceptionV3} & \textbf{0.9052} & \textbf{0.9001} & \textbf{0.9716} & \textbf{0.9345} \\
        EfficientNet B7 & 0.7146 & 0.7648 & 0.8518 & 0.8059 \\
        RadImagenet - DenseNet101 & 0.7746 & 0.8424 & 0.8892 & 0.8650 \\
        ResNet101 & 0.8146 & 0.9024 & 0.9492 & 0.9252 \\
        \hline
    \end{tabular}
    \caption{Performance Metrics of Various Models on the PCOS Dataset}
    \label{tab2:comparison}
\end{table}

The performance metrics of various models (Table~\ref{tab2:comparison}) used in this study highlight the effectiveness of the InceptionV3 model for classifying ultrasound images related to PCOS diagnosis. InceptionV3 achieved an accuracy of \textbf{90.52\%}, with high precision (\textbf{90.01\%}) and recall (\textbf{97.16\%}), resulting in a robust F1 score of \textbf{93.45\%} that indicates its strong capability in distinguishing between healthy and unhealthy images. In comparison, ResNet101 also demonstrated commendable performance with an accuracy of \textbf{81.46\%}, while EfficientNet B7 and RadImagenet - DenseNet101 showed lower accuracies of \textbf{71.46\%} and \textbf{77.46\%}, respectively, indicating that InceptionV3 outperformed all other models in this classification task, making it the most reliable choice for this application.

\subsection{Sample Predictions with LIME Explanations}

Sample predictions from the validation and test dataset are shown below along with their LIME explanations:

\begin{figure}[htbp]
    \centering
        \includegraphics[width=0.8\linewidth]{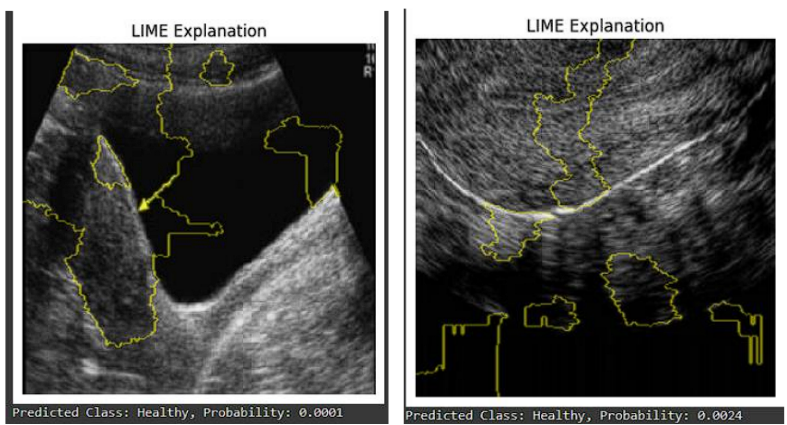}
        \caption{Comparison of LIME explanations for a correctly classified validation image (left) and test image (right). In both instances, the model accurately predicted Healthy, with LIME highlighting key regions which model focus on.}
    \label{fig4:LIME_Exp}
\end{figure}

\begin{figure}[htbp]
    \centering
        \includegraphics[width=0.8\linewidth]{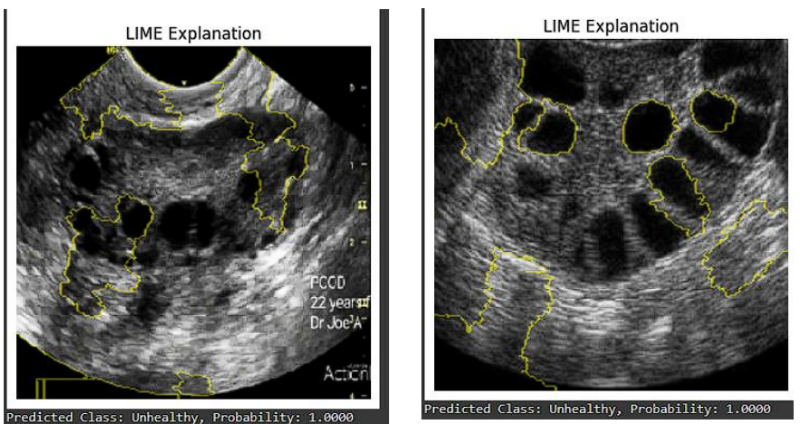}
        \caption{Comparison of LIME explanations for a correctly classified validation image (left) and test image (right). In both instances, the model accurately predicted Unhealthy, with LIME highlighting key regions which model focus on.}
    \label{fig5:LIME}
\end{figure}

This approach provides valuable insights into how the model arrives at its predictions, enhancing interpretability and trust in its decisions. By visualizing the contributions of different parts of the image, healthcare professionals can better understand the rationale behind the model's classifications. This transparency is particularly important in medical applications, where understanding the basis for a diagnosis can guide further clinical decision-making.

The use of LIME helps to demystify the model's behavior, allowing users to see which features were pivotal in determining whether an ultrasound image was classified as healthy or unhealthy. This not only aids in validating the model's performance but also fosters confidence among practitioners who may rely on automated systems for diagnostic support. Overall, integrating LIME explanations into the prediction process enhances the interpretability of deep learning models in medical imaging contexts, ensuring that they are not only accurate but also comprehensible to end-users.

\subsection{Computational Efficiency}

The computational efficiency of the InceptionV3 model was assessed based on training time and inference time during the classification task. The model was trained on an Nvidia GTX 1650 GPU with 8 GB of RAM, completing the training process for a total of 50 epochs in approximately 3 hours. This training duration reflects the model's ability to learn effectively from the dataset while utilizing available computational resources efficiently.
In terms of inference, the average time taken to classify a single image through the entire pipeline was around 30 seconds. This includes all preprocessing steps, prediction, and any necessary post-processing. Understanding these efficiency metrics is crucial for evaluating the practicality of deploying the model in real-world applications, particularly in clinical settings where timely decision-making is essential. The relatively quick inference time allows for effective integration into diagnostic workflows, enhancing the potential for real-time analysis of ultrasound images related to PCOS diagnosis.

\section{Discussion}\label{sec4}

Our team achieved a 3rd place ranking in the AUTO PCOS challenge, demonstrating the effectiveness of our deep learning approach for classifying ultrasound images related to PCOS diagnosis. This achievement underscores the potential of leveraging pre-trained convolutional neural networks, specifically InceptionV3, for this critical healthcare application.

Our methodology, as detailed in the Methods section, involved a structured pipeline encompassing image preprocessing, careful data preparation including a robust training/validation/test split, transfer learning using InceptionV3 with custom classification layers, and rigorous model evaluation. The choice of InceptionV3 was driven by extensive experimentation with several other architectures, including EfficientNet B7, RadImagenet-DenseNet101, ResNet101, and others (as shown in Table \ref{tab2:comparison}). While these models demonstrated varying degrees of success, InceptionV3 consistently outperformed them on our validation set, exhibiting a superior balance of precision and recall, ultimately leading to the highest F1-score and accuracy. This highlights the importance of empirical model selection, as even state-of-the-art architectures may not generalize equally well to specific datasets and tasks.

The performance metrics achieved by our InceptionV3 model on the validation set (Table \ref{tab1:performance}) are compelling. The high accuracy (90.52\%) indicates a strong overall ability to distinguish between healthy and unhealthy ultrasound images. The excellent recall (97.16\%) is particularly noteworthy in a medical context, as it signifies the model's high sensitivity in identifying true positive (unhealthy) cases, minimizing the risk of missed diagnoses. The precision (90.01\%) ensures that when the model predicts a case as unhealthy, it is highly likely to be correct, reducing unnecessary follow-up procedures and patient anxiety. The F1-score (93.45\%), which balances precision and recall, further confirms the robustness of our model.

The inclusion of LIME explanations (Figures \ref{fig4:LIME_Exp} and \ref{fig5:LIME}) provides crucial insights into the model's decision-making process. By visualizing the image regions that most strongly influenced the predictions, we gain a level of transparency that is essential in medical applications. This interpretability allows clinicians to understand *why* the model made a particular prediction, fostering trust and facilitating integration into clinical workflows. While we have provided examples, a more comprehensive analysis of LIME explanations across a larger and more diverse test set would further strengthen the understanding of the model's behavior.

Our evaluation also considered computational efficiency. The training time of approximately 3 hours on a consumer-grade GPU (Nvidia GTX 1650) is reasonable and demonstrates the feasibility of training such models with readily available hardware. The inference time of around 30 seconds per image, while not real-time, is acceptable for many diagnostic scenarios, especially when considering the potential for batch processing. Future work could explore optimizations to reduce inference time further for real-time applications.

Despite the strong performance, several limitations should be acknowledged. The dataset size, while sufficient for the challenge, could be expanded to improve the model's generalization capabilities and robustness to variations in image quality, ultrasound equipment, and patient demographics. Further investigation into the impact of different preprocessing techniques and hyperparameter tuning could also potentially yield further performance gains. Additionally, while LIME provides valuable local explanations, exploring other interpretability techniques could offer complementary insights.

In conclusion, our 3rd place ranking in the AUTO PCOS challenge demonstrates the effectiveness of our deep learning-based approach using InceptionV3 for classifying ultrasound images for PCOS diagnosis. The model's high performance metrics, coupled with the interpretability provided by LIME explanations, suggest its potential as a valuable tool for assisting healthcare professionals in clinical practice. Future work will focus on addressing the identified limitations to further enhance the model's performance, robustness, and clinical applicability.

\section{Conclusion}\label{sec5}

In conclusion, our 3rd place ranking in the AUTO PCOS challenge validates the efficacy of our deep learning approach for classifying ultrasound images for PCOS diagnosis. By leveraging the pre-trained InceptionV3 architecture within a carefully designed pipeline encompassing preprocessing, data preparation, targeted training, and rigorous evaluation, we achieved compelling performance metrics. The model demonstrated a high degree of accuracy (90.52\%), exceptional recall (97.16\%), and robust precision (90.01\%), culminating in a strong F1-score of 93.45\%. This performance, coupled with the valuable insights provided by LIME explanations, underscores the potential of our approach to assist healthcare professionals in clinical practice. The comparative analysis against other architectures further solidified the selection of InceptionV3 as the most suitable model for this specific task, highlighting the importance of thorough model experimentation. While acknowledging limitations related to dataset size and potential for further optimization, our work contributes a robust and interpretable deep learning solution for PCOS diagnosis using ultrasound imagery. This achievement not only demonstrates the power of transfer learning in medical image analysis but also paves the way for future research aimed at refining diagnostic tools and ultimately improving patient care in the context of PCOS. Our findings suggest that with further refinement and validation on larger, more diverse datasets, this approach could be effectively translated into a clinically relevant tool for improved and more accessible PCOS diagnosis.

\section{Acknowledgments}\label{sec6}
As participants in the Auto-PCOS Classification Challenge, we fully comply with the competition's rules as outlined in \cite{hub2024auto} and the challenge website. Our methods are based on the train and test datasets provided in the official release in \cite{handaPCOStraining} and \cite{handaPCOStesting}.

\end{document}